\documentclass[10pt, conference, letterpaper]{IEEEtran}

\usepackage{tabularx, booktabs}
\usepackage{amsmath,amssymb,amsfonts}
\usepackage{amssymb}
\usepackage{xcolor}
\usepackage{listings}
\usepackage[T1]{fontenc}
\usepackage{algorithm2e}
\usepackage[caption=false]{subfig}

\usepackage{tikz}
\usetikzlibrary{positioning, fit}
\usetikzlibrary{shapes}

\usepackage{pgfplots}
\pgfplotsset{compat=newest}
\usepgfplotslibrary{statistics}

\author{\IEEEauthorblockN{Mohammad Riftadi}\IEEEauthorblockA{Delft University of Technology\\M.Riftadi@student.tudelft.nl} \and
\IEEEauthorblockN{Jorik Oostenbrink}\IEEEauthorblockA{Delft University of Technology\\J.Oostenbrink@tudelft.nl} \and \IEEEauthorblockN{Fernando Kuipers}\IEEEauthorblockA{Delft University of Technology\\F.A.Kuipers@tudelft.nl}}

\begin{document}
\title{GP4P4: Enabling Self-Programming Networks}

\maketitle

\begin{abstract}
	Recent advances in programmable switches have enabled network operators to build high-speed customized network functions. Although this is an important step towards self-* networks, operators are now faced with the burden of learning a new language and maintaining a repository of network function code. Inspired by the Intent-Based Networking paradigm, we propose a new framework, GP4P4: a genetic programming approach able to autonomously generate programs for P4-programmable switches directly from network intents. We demonstrate that GP4P4 is able to generate  various small network functions in up to a few minutes; an important first step towards realizing the vision of `Self-Driving' networks.
\end{abstract}

\section{Introduction}
\label{section:intro}
The concept of Self-Driving Networks, analogous to the concept of Self-Driving Cars, has been a Utopian dream in the field of computer networks. That ultimate goal of running a network that behaves solely based on our intent is rapidly coming in reach through fast advances in the domains of network programmability and artificial intelligence \cite{kellerer2019adaptable, boutaba2018comprehensive}.

The introduction of the network programming language P4 \cite{bosshart2014p4}, which allows for data-plane programmability, has enabled network operators to construct high-speed network functions customized to their own needs. However, this does require them to create and maintain a large library of network function code, which is prone to human error. Moreover, the move from P4$_{14}$ \cite{P414lang2018} to P4$_{16}$ \cite{P416lang2017} introduced major code-breaking changes to the language. As languages keep evolving, to remain up to date, a network operator would need to adjust his entire catalog of P4 programs. 

In \cite{Riftadi2019}, the authors proposed an intent-based programming framework for P4, which can automatically create and install P4 programs using a library of P4 templates. Although this is a step in the right direction and simplifies the process of changing network functionality on the fly, it shifts the problem of maintaining a catalog of P4 programs to maintaining a catalog of P4 templates.

To avoid these problems entirely, we propose to leave the programming of the network to the network itself by enabling it to automatically generate data-plane code based on sets of less complex, human-readable rules or intents. These programs can then be used as templates to create larger, more complex programs or directly installed in the network. As a proof of concept, we present GP4P4, a framework that allows operators to modify their network functionality near instantaneously without modifying any code themselves. We believe this framework is an important first step towards a future where \emph{self-programming} networks can fully program and adapt themselves to their current goals and circumstances with minimum intervention by network operators.

Machine-learning has recently been applied within the control-plane (see \cite{xie2018survey}) and to boost the performance of network functions (e.g., \cite{8262813, geng2019simon, yang2018empowering}), but the utilization of machine learning techniques to generate network functions themselves has yet to be considered. Also, a few position papers on self-driving networks have appeared \cite{feamster2017and, juniper, kalmbach2018empowering, yaqoob2018analyzing, 8262813}, but again a concrete framework that enables the network to program itself is missing.

Our main contributions are: (1) GP4P4 itself, a framework for automatically generating P4 programs using techniques adapted from Linear Genetic Programming (LGP). LGP is a machine-learning technique to ``evolve'' an initially randomized population of programs towards satisfying an objective function~\cite{Brameier2010}; (2) An evaluation module required to make LGP suitable for dataplane programmability. Our proposed evaluation module evaluates programs by creating synthetic network traces and simulating the output of P4 programs on these traces. In this regard, GP4P4 is fully self-sufficient and does not depend on any external network traces or physical switches; (3) Proof-of-Concept experiments demonstrating the efficacy of GP4P4.

\section{GP4P4}
\begin{figure*}
\centering
\includegraphics[width=0.9\textwidth]{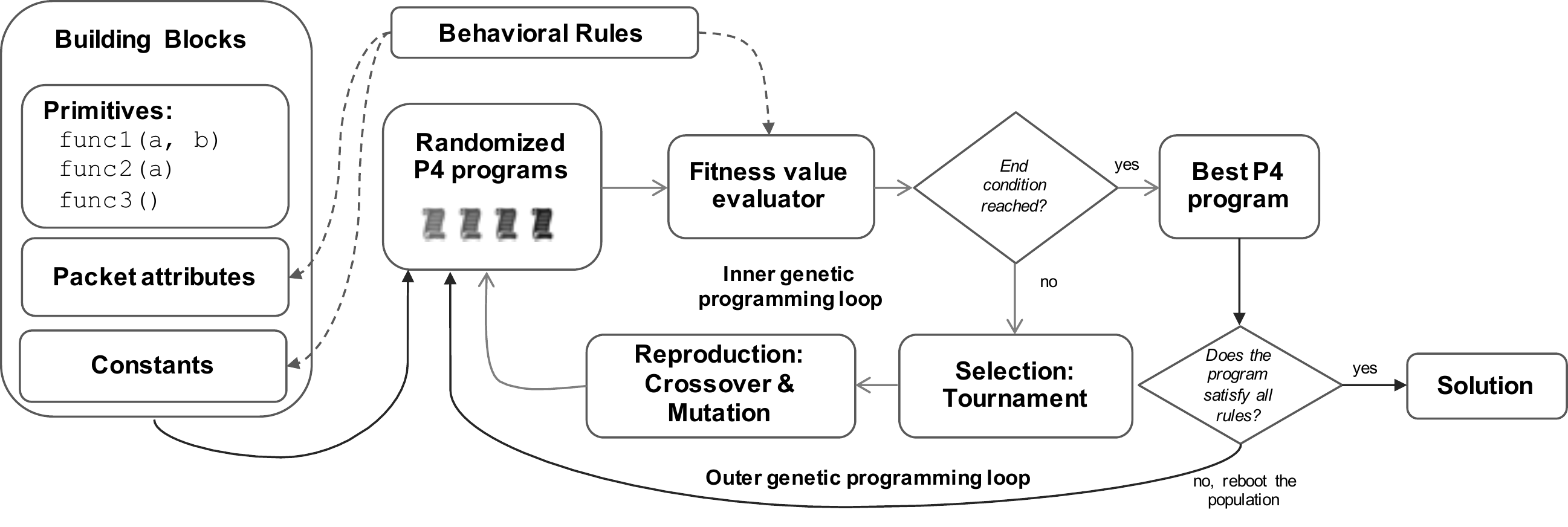}
\caption{Program generation overview.}
\label{fig:gp_overview}
\end{figure*}
Figure \ref{fig:gp_overview} gives a high-level overview of GP4P4. Behavioral rules -- the intents of the network operator -- lie at the base of our framework. They are analyzed to obtain the P4 building blocks that the framework uses to create P4 programs, as well as for evaluating the programs during and after their generation. In the inner loop, GP4P4 evolves programs using LGP. If the best of these programs satisfies all behavioral rules, GP4P4 presents this program as the solution. If not, it reboots the population of P4 programs and restarts the inner loop. This process continues until a solution has been found.

\subsection{Behavioral Rules}
We describe combinations of network functions as a set of \emph{behavioral rules} on packet attributes (headers and metadata). If these rules are followed for each packet, we say that the P4 program is \emph{valid}. In contrast to P4 programs themselves, these rules describe the intended outcome of a program, and not its methodology. In other words, these rules can be seen as a low-level description of the intents of a network operator.

In GP4P4, behavioral rules are expressed in the form of IF-THEN predicates connecting packet input conditions to output conditions. For any packet for which the input conditions (IF) are true, we require that the output conditions (THEN) are also true. Both the input and output conditions are expressed as one or more equal (EQ) or not equal (NEQ) Boolean expressions on packet attributes combined with AND and OR. Packet attributes are referenced using dot notation (e.g. \verb|pkt_in.src_ip|). For example, the rule 
\begin{verbatim}
IF (pkt_in.src_ip EQ 192.168.1.1) 
THEN (pkt_out.out_port EQ 2)
\end{verbatim}
means that if the source IP address of an incoming packet is 192.168.1.1, the packet should be outputted from port 2. As a default, GP4P4 adds rules to ensure any unmatched packet attributes are kept constant.

We see these rules as a type of low-level intents, in between natural language high-level intents and P4 code itself. Thus, although they can be easily created by network operators themselves, behavioral rules could also be generated by natural language processors as an intermediary step between natural language and P4 code, allowing these processors to generate P4 code through GP4P4.

\subsubsection{NAT Example}
\label{sec:NAT_example}
\begin{figure}[t!]
\centering
\includegraphics[width=\columnwidth]{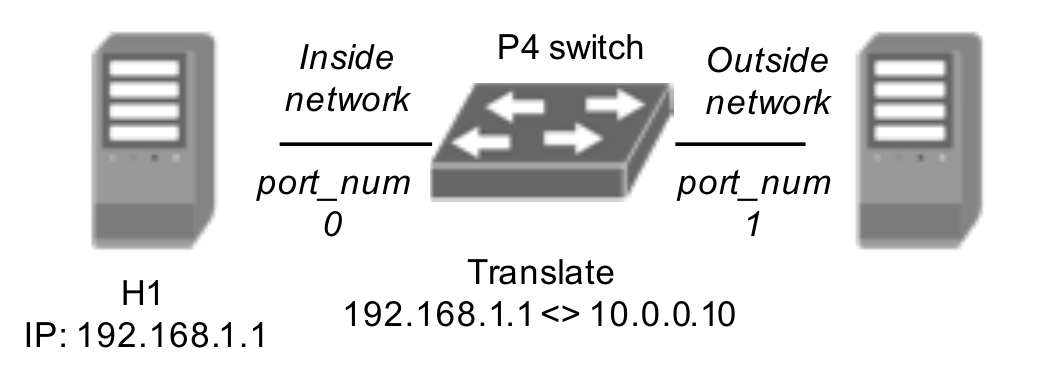}
\caption{NAT example topology.}
\label{fig:nat_topology}
\end{figure}

Consider the example situation in Figure \ref{fig:nat_topology}. Switch S1 connects two networks, \textit{inside} and \textit{outside}. We want to install a ``static source Network Address Translation (NAT)'' network function on the switch that automatically replaces the source destination IP address of host H1 in \textit{inside}, 192.168.1.1, with 10.0.0.10 in any outgoing packet, and the destination IP address 10.0.0.10 with 192.168.1.1 in any incoming packet. The IP addresses of all other packets should be left unchanged. To determine if packets are from inside or outside, we can match on their input port (\verb|pkt_in.port_num|); packets arriving at port 0 are moving from inside to outside, and packets arriving at port 1 are moving from outside to inside. Figure \ref{fig:nat_rules} shows the resulting 2 behavioral rules for this network function.

\begin{figure}[t!]
\centering
\begin{verbatim}
RULE1: 
IF   (pkt_in.port_num EQ 0
      AND pkt_in.src_ip EQ 192.168.1.1)
THEN (pkt_out.src_ip EQ 10.0.0.10)
RULE2: 
IF   (pkt_in.port_num EQ 1
      AND pkt_in.dst_ip EQ 10.0.0.10)
THEN (pkt_out.dst_ip EQ 192.168.1.1)
\end{verbatim}
\caption{NAT example behavioral rules.}
\label{fig:nat_rules}
\end{figure}

\subsection{Program Generation}
\label{sec:program_generation}

In Genetic Programming (GP), an initially randomized population of programs is gradually evolved to satisfy an objective function by selection and reproduction, similarly to the biological concept of natural selection. To move through the search space, reproduced programs are randomly modified by \emph{mutation} and/or \emph{crossover} operations.

An important concept in GP is that of the phenotype versus the genotype. The phenotype is the program itself, while the genotype is an internal lower-level representation of the program that is more suitable for GP. Given a genotype, we can directly construct the phenotype by de-encoding this representation. In practice, GP has three major genotype representations: (1) linear, (2) tree-based, and (3) graph-based. In a tree-based approach, programs permanently branch of after every IF-statement. Thus, this representation is more likely to evolve nested IF-statements than successive IF-statements. The graph-based approach does not suffer from this ``problem,'' but limits our ability to perform crossover\footnote{Combining information from two parent programs to create new offspring.}. Crossover is vital for creating programs that satisfy our behavioral rules, as it allows a program that satisfies one rule to ``merge'' with a program that satisfies another rule to, hopefully, create an offspring that satisfies both rules. As we want to be able to evolve programs with both nested and successive IF-statements, as well as make use of crossover, we use Linear Genetic Programming (LGP) in GP4P4.
LGP evolves sequences of \emph{primitives} of an imperative programming language. Each of these primitives represents a snippet of code in the phenotype program. In GP4P4 each primitive corresponds to a basic one-line declaration in P4, such as \verb|src_ip = 10.0.0.10;|. We assume a set of primitives is provided to GP4P4 every time a new program has to be generated. 

As depicted in Figure~\ref{fig:gp_overview}, program generation in GP4P4 runs in two loops: the outer and inner loop. In the inner loop, LGP is applied to evolve an initial random population of programs into a program that satisfies all behavioral rules. The inner loop finishes either when a solution has been found or when this process takes too long. In the later case, the outer loop restarts the inner loop with a new initial population. This process helps prevent the program generation module from getting stuck in a sub-optimal local minimum. 

If the number of maximum inner loop iterations is too low, GP4P4 might be unable to generator some, more difficult, programs in reasonable time, as the generation process is interrupted before the program can be completed. Conversely, if the maximum number of inner loop iterations is too high, GP4P4 takes too long to restart a stuck inner loop. The optimal number of maximum inner loop iterations differs on a case by case basis. Thus, GP4P4 starts with a small maximum number of inner loop iterations of $\text{init\_it}$ and doubles the number of allowed iterations every time the loop is restarted, until it reaches $\text{max\_it}$. This process significantly reduces the total program generation time of GP4P4 compared to choosing a fixed number of inner loop iterations.

\subsubsection{Building Blocks}
\label{sec:building_blocks}

A primitive may read (write) from (to) any switch register, metadata value, or packet header. Additionally, a primitive may also access one or more constants (e.g. port \emph{0}). In GP4P4, we treat all these input/output locations and values as \emph{registers}. We store these registers in a single array, and allow a primitive to operate on any combination of registers in this array. Note that this means that the same primitive may correspond to different P4 declarations, depending on which registers it accesses. The registers of constants are read-only and are not allowed to be written to. We do not have to store the explicit values of each register, but only to which part of the memory or packet they refer. We refer to the combination of registers and primitives as the \emph{building blocks} of a GP4P4 program.

To construct GP4P4 primitives, P4 declarations are simplified and written in prefix notation. For example, the P4 declaration \verb|var3 = var5;| is transformed to the GP4P4 primitive \verb|ASSIGN(var3, var5)|. Although P4 declarations are normally written in infix notation, it is easier to encode genes in prefix format. The if-then statement (\verb|if() { }|) is cut into two primitives, corresponding to \verb|if() {| and \verb|}|. In addition, we restrict these primitives to two input registers, and create a separate primitive for each possible comparison operator: \verb|IF_EQ(a,b)| for \verb|if (a == b) {|,  \verb|IF_NEQ(a,b)| for \verb|if (a != b) {|, and \verb|ENDIF| for \verb|}|. We do not allow any other control-flow statements, such as the else statement. Although this choice of primitives is rather limited, it still supports a wide range of possible declarations, albeit in the form of multiple primitives. For example, \verb|if(a == b && b == c) {| is represented as \verb|IF_EQ(a,b), IF_EQ(b,c)|.  

P4 allows for a wide array of possible memory locations, metadata values and packet headers. Including all these possibilities as registers would severely hamper the ability of GP4P4 to evolve programs into the right direction. Thus, GP4P4 automatically extracts all registers from the behavioral rules themselves. A packet attribute or constant is included as a register if and only if it is used in at least one of the behavioral rules.

As an optimization step, we categorize each attribute and constant by its data type, such as integer, string, Boolean, or IP address. We then limit the registers each primitive is allowed to access by type. This reduces the search space and ensures each primitive + register combination translates to correct P4 code.

\subsubsection{Initial Population}
\label{sec:init_pop}
To initialize the inner loop, we generate a population of $N$ syntactically correct programs of primitives with a length between min\_len and max\_len. To construct each program, GP4P4 first randomly picks a program length between min\_len and max\_len. Then, it randomly generates this number of primitives, randomly selects valid registers for each primitive, and puts the primitives in sequence. This process is repeated every time the inner loop is restarted.

\subsubsection{Selection and Reproduction}
Within each iteration of the inner loop, GP4P4 holds two tournaments between $t_r \times N$ randomly selected programs, where $t_r$ is the tournament size ratio. The program with the highest fitness value of each tournament (or winner) is chosen for reproduction, while the bottom $n_r$ programs (or losers) of each tournament are chosen to be replaced by the offsprings of the two winners. In LGP, the new set of programs created by replacing the losers by the offspring of the winners is called a new \emph{generation}. The inner loop continues the process of generating new generations until it either finds a valid program or reaches a pre-set generation limit.  

Each of the $2 \times n_r$ offspring is created in pairs of two:
\begin{enumerate}
    \item Duplicate both winners
    \item Perform a crossover between both offspring with probability $P_c$
    \item Mutate offspring 1 with probability $P_m$
    \item Mutate offspring 2 with probability $P_m$
\end{enumerate}
We compute and store the fitness value of each program as soon as it is created. This way, we reduce the number of fitness values that need to be computed every iteration from $t_r \times N$ to $2 \times n_r$.

\subsubsection{Mutation}
To mutate a program, GP4P4 first selects a random index $i$ in the program. Then, with equal probability, it either adds a new random primitive to the program at $i+1$, removes the current primitive at $i$, or replaces the current primitive at $i$ with a new random primitive. Random primitives are generated in the same way as described previously in Section ~\ref{sec:init_pop}, with a few notable exceptions: To help evolve the program towards satisfying new rules, we generate new random if-then primitives with a higher probability than other primitives. GP4P4 selects a new, random if-then primitive with probability $P_\text{if}$ and a non-if-then primitive with probability $1 - P_\text{if}$. In addition, to prevent new if-then primitives from dropping the fitness level of the program, GP4P4 adds an \verb|ENDIF()| primitive directly after every new if-then primitive it adds to a program. Similarly, when removing an if-then or \verb|ENDIF()| primitive, GP4P4 also removes the corresponding \verb|ENDIF()| or if-then primitive. 

\subsubsection{Crossover}
To perform a crossover between two programs, GP4P4 randomly selects a \emph{unit} of code of both programs and swaps these units with each other. In GP4P4, a unit is either a single non-if-then primitive or a sequence of primitives starting with an if-then primitive and ending with its corresponding \verb|ENDIF()| primitive. By only swapping valid blocks of code, we ensure that the resulting two programs remain syntactically valid. 

\subsection{Program Evaluation}
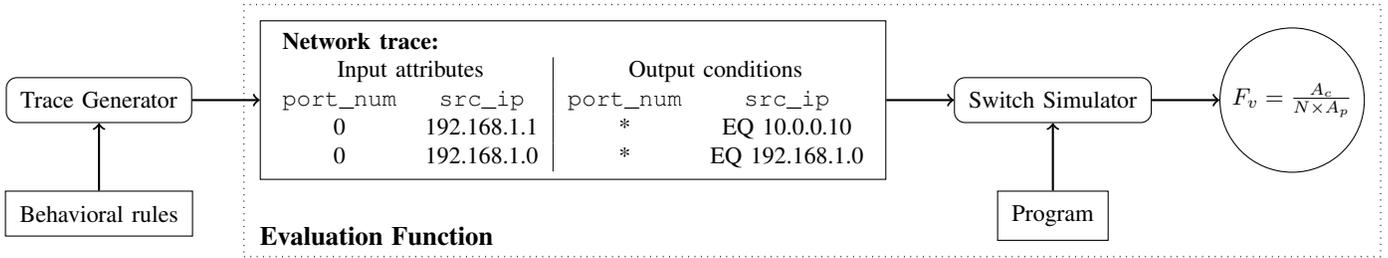
\begin{figure*}[t!]
    \centering
    \begin{tikzpicture}[transform shape, scale = 0.9]
    \node [draw, inner sep = 6] (Rules) {Behavioral rules};
    \node [draw, inner sep = 6, rounded corners, above = of Rules] (Generator) {Trace Generator};
    
    \node [draw, right = of Generator] (Trace) {\begin{tabular}{cc|cc}
    \multicolumn{4}{l}{\bfseries Network trace:}\\
    \multicolumn{2}{c|}{Input attributes} & \multicolumn{2}{c}{Output conditions} \\
    \verb|port_num| & \verb|src_ip| & \verb|port_num| & \verb|src_ip| \\
    0 & 192.168.1.1 & * & EQ 10.0.0.10 \\
    0 & 192.168.1.0 & * & EQ 192.168.1.0
    \end{tabular}};
    
    \node [draw, inner sep = 6, rounded corners, right = of Trace] (Simulator) {Switch Simulator};
    \node [draw, inner sep = 6, below = of Simulator] (Program) {Program};
    
    \node [draw, circle, right = of Simulator] (Fitness) {$F_v = \frac{A_c}{N \times A_p}$};
    
    \path [draw, thick, ->] (Rules) edge (Generator);
    \path [draw, thick, ->] (Generator) edge (Trace);
    \path [draw, thick, ->] (Trace) edge (Simulator);
    \path [draw, thick, ->] (Simulator) edge (Fitness);
    \path [draw, thick, ->] (Program) edge (Simulator);
    
    \node [draw, transform shape=false, dotted, fit = (Trace) (Program) (Fitness), inner sep = 6, align = left, text height = 85] (EvFunc) {\textbf{Evaluation Function}};
    \end{tikzpicture}
    \caption{Evaluation module overview.}
    \label{fig:evaluation_overview}
\end{figure*}

The evaluation module plays a critical role in GP4P4, as it guides the evolution of programs in the right direction, as well as checks if a program satisfies all behavioral rules. A good evaluation function should evaluate, in fine granularity, how close a program is to satisfying all rules and express this in a numerical value. In the case of P4 programs, this is not a straightforward process, as programs may seemingly satisfy a rule for one packet, while breaking it for another. Figure \ref{fig:evaluation_overview} gives an overview of the evaluation module.

First, the \emph{Trace Generator} generates a synthetic network trace of packets and output conditions based on the behavior rules supplied to the framework. Then, the \emph{Switch Simulator} simulates the program and processes the network trace. For each packet in the trace, the simulator counts the number of packet output attributes that satisfy the behavioral rules. Finally, the total number of these valid output attributes over all packets in the trace, $A_c$, is normalized to obtain the fitness value, $F_v$, by dividing it by the total number of packets in the network trace, $N$, times the number of output attributes per packet, $A_p$: $F_v := \frac{A_c}{N \times A_p}$.

\subsubsection{Trace Generator}
If a behavioral rule contains multiple expressions combined with OR, it is important that the final program is valid for all possible cases. Thus, the Trace Generator first splits the IF statement of each behavioral rule into its disjunctive normal form and creates a separate rule for each of its clauses. In addition, to ensure the program does not modify any attributes if it does not match any rules, the Trace Generator also adds the complement of all behavioral rules as a default rule.

For each of these created rules, the Trace Generator creates $k$ packets. Packet input attributes are created in a semi-randomized fashion to match the IF conditions of the rule, while the output attribute conditions are directly taken from the THEN conditions of any rule that match the randomly-created packet. By creating packets for each rule, we ensure that the fitness evaluation function evaluates programs on each rule as well. To reduce computation time, the same network trace is re-used throughout the inner and outer genetic programming loops. Thus, the Trace Generator is only run once, just before starting the outer genetic programming loop.

\subsubsection{Switch Simulator}
Compiling a program to P4, and then running the program on a real or emulated switch can take a significant amount of time. We propose running and evaluating each program on a simulated switch instead,  while guaranteeing the same output/fitness as a real switch.

To save time, the simulator (written in Python) runs directly on the sequence of primitives (the genotype) described in Section ~\ref{sec:program_generation} instead of on P4 code (phenotype). When simulating a program, the Switch Simulator first initializes a new list of registers, as described in Section ~\ref{sec:building_blocks}. It then ``runs'' the program on each packet of the network trace by
\begin{enumerate}
    \item Copying the packet attributes to the corresponding registers.
    \item Interpreting the GP4P4 primitives line by line, reading and modifying the register values whenever required.
    \item Copying the output packet attributes from the corresponding registers.
\end{enumerate}
The fitness value of the program is then determined by counting the total number of satisfied output conditions, $A_c$, and dividing this value by the total number of output conditions, $N \times A_p$.

In the Switch Simulator, all primitives are assigned their own Python function. Consequently, to interpret a GP4P4 primitive, the simulator simply executes the corresponding Python function. If-then primitives form their own special case: when the simulator encounters an if-then primitive, it checks if its condition is true. If it is, the simulator continues to the next line. If not, the simulator searches for and skips forward to the corresponding \verb|ENDIF()| primitive. To prevent the simulator from jumping to the end of a nested if-then block instead, it keeps track of its current depth while searching for the correct \verb|ENDIF()| primitive.

\section{Experiments}
\label{section:experiments}
\begin{table}[t!]
\centering
 \caption{Demonstration network functions}
\label{tab:nf}
\begin{tabular}{ccc}
\bfseries Network Function & \bfseries rules & \bfseries primitives \\ 
Network Adress Translation (NAT) & 2 & 3\\
Firewall & 1 & 3\\
Server Balancer & 2 & 3\\
Link Balancer & 2 & 3\\
DSCP Marker & 2 & 3\\
Router & 2 & 4\\
Port Address Translation (PAT) & 1 & 3\\
\end{tabular}
\end{table}

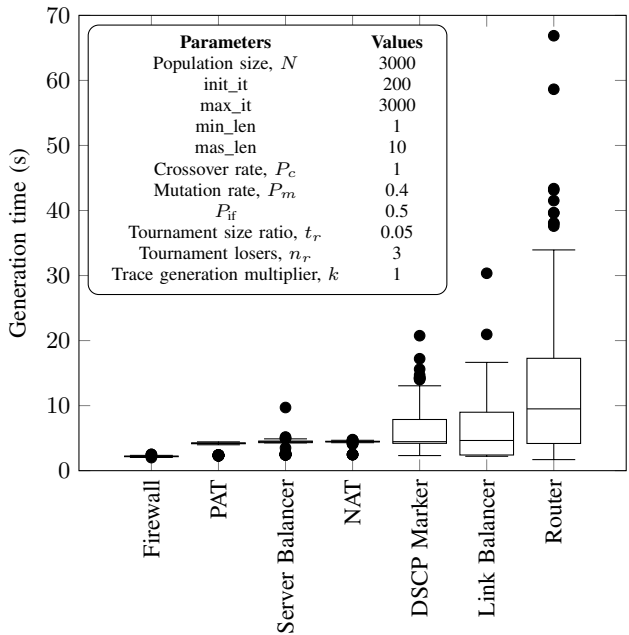
\begin{figure}[t!]
    \centering
    \begin{tikzpicture}\small
    \begin{axis}[
        width = \linewidth,
        ymin = 0,
        ymax = 70,
        ytick distance = 10,
        ylabel = Generation time (s),
        boxplot/draw direction=y,
        xtick={1,2,3,4,5,6,7},
        xticklabels={Firewall, PAT, Server Balancer, NAT, DSCP Marker, Link Balancer, Router},
        x tick label style={rotate = 90}
        ]
        \addplot [boxplot] table[y index = 0]{experiment_results/default_firewall.dat};
        \addplot [boxplot] table[y index = 0]{experiment_results/default_pat.dat};
        \addplot [boxplot] table[y index = 0]{experiment_results/default_slb.dat};
        \addplot [boxplot] table[y index = 0]{experiment_results/default_nat.dat};
        \addplot [boxplot] table[y index = 0]{experiment_results/default_dscp_marker.dat};
        \addplot [boxplot] table[y index = 0]{experiment_results/default_wlb.dat};
        \addplot [boxplot] table[y index = 0]{experiment_results/default_router.dat};
        
        \node [draw, rounded corners, anchor = south west, fill = white] at (0.05, 27) (insert) {\scriptsize
        \begin{tabular}{cc}
        \bfseries Parameters            & \bfseries Values \\
        Population size, $N$            & 3000\\
        init\_it     & 200 \\
        max\_it      & 3000 \\
        min\_len                        & 1    \\
        mas\_len                        & 10   \\
        Crossover rate, $P_c$    & 1 \\
        Mutation rate, $P_m$     & 0.4 \\
        $P_\text{if}$                   & 0.5 \\
        Tournament size ratio, $t_r$    & 0.05 \\
        Tournament losers, $n_r$        & 3 \\
        Trace generation multiplier, $k$& 1
\end{tabular}
        }; 
    \end{axis}
    \end{tikzpicture}
    \caption{Tukey boxplots of the generation times of 7 network functions.  Each network function was generated 100 times.}
    \label{fig:gen_time}
\end{figure}
\begin{figure*}[t!]
    \subfloat{%
    \begin{tikzpicture}\small
    \begin{axis}[
        width = 0.36\linewidth,
        ymin = 0,
        ymax = 15,
        ylabel = Generation time (s),
        y label style={yshift=0.5em},
        boxplot/draw direction=y,
        xtick={1,2,3,4,5},
        xticklabels={,,,,}
        ]
        \addplot [boxplot] table[y index = 0]{experiment_results/pop_firewall_1000.dat};
        \addplot [boxplot] table[y index = 0]{experiment_results/default_firewall.dat};
        \addplot [boxplot] table[y index = 0]{experiment_results/pop_firewall_5000.dat};
        \addplot [boxplot] table[y index = 0]{experiment_results/pop_firewall_7000.dat};
        \addplot [boxplot] table[y index = 0]{experiment_results/pop_firewall_9000.dat};
    \end{axis}
    \end{tikzpicture}}\hfill
    \subfloat{%
    \begin{tikzpicture}\small
    \begin{axis}[
        width = 0.36\linewidth,
        ymin = 0,
        ymax = 15,
        boxplot/draw direction=y,
        xtick={1,2,3,4,5},
        xticklabels={,,,,}
        ]
        \addplot [boxplot] table[y index = 0]{experiment_results/tsr_firewall_0.10.dat};
        \addplot [boxplot] table[y index = 0]{experiment_results/tsr_firewall_0.20.dat};
        \addplot [boxplot] table[y index = 0]{experiment_results/tsr_firewall_0.30.dat};
        \addplot [boxplot] table[y index = 0]{experiment_results/tsr_firewall_0.40.dat};
        \addplot [boxplot] table[y index = 0]{experiment_results/tsr_firewall_0.50.dat};
    \end{axis}
    \end{tikzpicture}}\hfill
    \subfloat{%
    \begin{tikzpicture}\small
    \begin{axis}[
        width = 0.36\linewidth,
        ymin = 0,
        ymax = 70,
        boxplot/draw direction=y,
        xtick={1,2,3,4,5},
        xticklabels={,,,,}
        ]
        \addplot [boxplot] table[y index = 0]{experiment_results/default_firewall.dat};
        \addplot [boxplot] table[y index = 0]{experiment_results/max_len_firewall_20.dat};
        \addplot [boxplot] table[y index = 0]{experiment_results/max_len_firewall_30.dat};
        \addplot [boxplot] table[y index = 0]{experiment_results/max_len_firewall_40.dat};
        \addplot [boxplot] table[y index = 0]{experiment_results/max_len_firewall_50.dat};
    \end{axis}
    \end{tikzpicture}}\\
    \subfloat{%
    \begin{tikzpicture}\small
    \begin{axis}[
        width = 0.36\linewidth,
        ymin = 0,
        ymax = 270,
        ylabel = Generation time (s),
        boxplot/draw direction=y,
        xtick={1,2,3,4,5},
        xticklabels={1000,,5000,,9000},
        xlabel = $N$
        ]
        \addplot [boxplot] table[y index = 0]{experiment_results/pop_router_1000.dat};
        \addplot [boxplot] table[y index = 0]{experiment_results/default_router.dat};
        \addplot [boxplot] table[y index = 0]{experiment_results/pop_router_5000.dat};
        \addplot [boxplot] table[y index = 0]{experiment_results/pop_router_7000.dat};
        \addplot [boxplot] table[y index = 0]{experiment_results/pop_router_9000.dat};
    \end{axis}
    \end{tikzpicture}}\hfill
    \subfloat{%
    \begin{tikzpicture}\small
    \begin{axis}[
        width = 0.36\linewidth,
        ymin = 0,
        ymax = 270,
        boxplot/draw direction=y,
        xtick={1,2,3,4,5},
        xticklabels={0.1,,0.3,,0.5},
        xlabel = $t_r$
        ]
        \addplot [boxplot] table[y index = 0]{experiment_results/tsr_router_0.10.dat};
        \addplot [boxplot] table[y index = 0]{experiment_results/tsr_router_0.20.dat};
        \addplot [boxplot] table[y index = 0]{experiment_results/tsr_router_0.30.dat};
        \addplot [boxplot] table[y index = 0]{experiment_results/tsr_router_0.40.dat};
        \addplot [boxplot] table[y index = 0]{experiment_results/tsr_router_0.50.dat};
    \end{axis}
    \end{tikzpicture}}\hfill
    \subfloat{%
    \begin{tikzpicture}\small
    \begin{axis}[
        width = 0.36\linewidth,
        ymin = 0,
        ymax = 480,
        boxplot/draw direction=y,
        xtick={1,2,3,4,5},
        xticklabels={10,,30,,50},
        xlabel = max\_len
        ]
        \addplot [boxplot] table[y index = 0]{experiment_results/default_router.dat};
        \addplot [boxplot] table[y index = 0]{experiment_results/max_len_router_20.dat};
        \addplot [boxplot] table[y index = 0]{experiment_results/max_len_router_30.dat};
        \addplot [boxplot] table[y index = 0]{experiment_results/max_len_router_40.dat};
        \addplot [boxplot] table[y index = 0]{experiment_results/max_len_router_50.dat};
    \end{axis}
    \end{tikzpicture}}
    \caption{Tukey boxplots of the generation times of Firewall (top row) and Router (bottom row) versus the population size ($N$), tournament size ratio ($t_r$), and maximum initial program length (max\_len).  All experiments were repeated 100 times.}
    \label{fig:parameters}
\end{figure*}
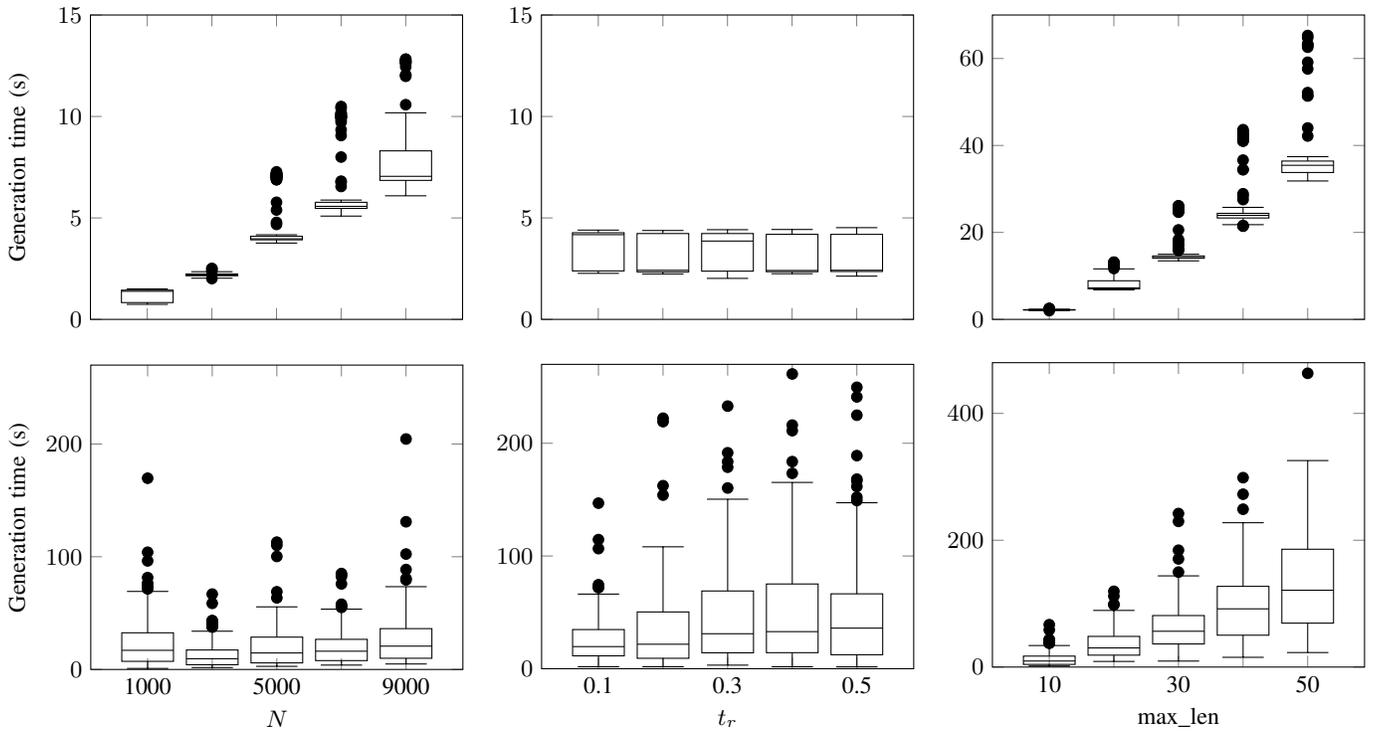

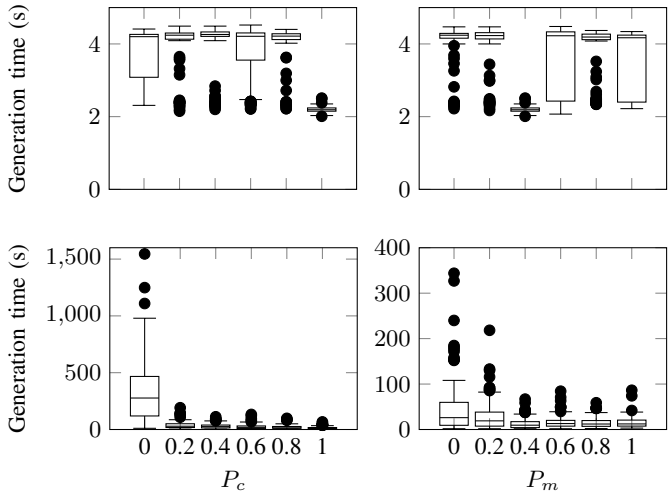
\begin{figure}[t!]
    \subfloat{%
    \begin{tikzpicture}\small
    \begin{axis}[
        width = 0.55\columnwidth,
        height = 4 cm,
        ymin = 0,
        ymax = 5,
        ylabel = Generation time (s),
        y label style={yshift=1.8em},
        boxplot/draw direction=y,
        xtick={1,2,3,4,5,6},
        xticklabels={,,,,}
        ]
        \addplot [boxplot] table[y index = 0]{experiment_results/crossover_firewall_0.00.dat};
        \addplot [boxplot] table[y index = 0]{experiment_results/crossover_firewall_0.20.dat};
        \addplot [boxplot] table[y index = 0]{experiment_results/crossover_firewall_0.40.dat};
        \addplot [boxplot] table[y index = 0]{experiment_results/crossover_firewall_0.60.dat};
        \addplot [boxplot] table[y index = 0]{experiment_results/crossover_firewall_0.80.dat};
        \addplot [boxplot] table[y index = 0]{experiment_results/default_firewall.dat};
    \end{axis}
    \end{tikzpicture}}\hfill
    \subfloat{%
    \begin{tikzpicture}\small
    \begin{axis}[
        width = 0.55\columnwidth,
        height = 4 cm,
        ymin = 0,
        ymax = 5,
        boxplot/draw direction=y,
        xtick={1,2,3,4,5,6},
        xticklabels={,,,,}
        ]
        \addplot [boxplot] table[y index = 0]{experiment_results/mutation_firewall_0.00.dat};
        \addplot [boxplot] table[y index = 0]{experiment_results/mutation_firewall_0.20.dat};
        \addplot [boxplot] table[y index = 0]{experiment_results/default_firewall.dat};
        \addplot [boxplot] table[y index = 0]{experiment_results/mutation_firewall_0.60.dat};
        \addplot [boxplot] table[y index = 0]{experiment_results/mutation_firewall_0.80.dat};
        \addplot [boxplot] table[y index = 0]{experiment_results/mutation_firewall_1.00.dat};
    \end{axis}
    \end{tikzpicture}}\\
    \subfloat{%
    \begin{tikzpicture}\small
    \begin{axis}[
        width = 0.55\columnwidth,
        height = 4 cm,
        ymin = 0,
        ymax = 1600,
        ylabel = Generation time (s),
        boxplot/draw direction=y,
        xtick={1,2,3,4,5,6},
        xticklabels={0,0.2,0.4,0.6,0.8,1},
        xlabel = $P_c$
        ]
        \addplot [boxplot] table[y index = 0]{experiment_results/crossover_router_0.00.dat};
        \addplot [boxplot] table[y index = 0]{experiment_results/crossover_router_0.20.dat};
        \addplot [boxplot] table[y index = 0]{experiment_results/crossover_router_0.40.dat};
        \addplot [boxplot] table[y index = 0]{experiment_results/crossover_router_0.60.dat};
        \addplot [boxplot] table[y index = 0]{experiment_results/crossover_router_0.80.dat};
        \addplot [boxplot] table[y index = 0]{experiment_results/default_router.dat};
    \end{axis}
    \end{tikzpicture}}\hfill
    \subfloat{%
    \begin{tikzpicture}\small
    \begin{axis}[
        width = 0.55\columnwidth,
        height = 4 cm,
        ymin = 0,
        ymax = 400,
        boxplot/draw direction=y,
        xtick={1,2,3,4,5,6},
        xticklabels={0,0.2,0.4,0.6,0.8,1},
        xlabel = $P_m$
        ]
        \addplot [boxplot] table[y index = 0]{experiment_results/mutation_router_0.00.dat};
        \addplot [boxplot] table[y index = 0]{experiment_results/mutation_router_0.20.dat};
        \addplot [boxplot] table[y index = 0]{experiment_results/default_router.dat};
        \addplot [boxplot] table[y index = 0]{experiment_results/mutation_router_0.60.dat};
        \addplot [boxplot] table[y index = 0]{experiment_results/mutation_router_0.80.dat};
        \addplot [boxplot] table[y index = 0]{experiment_results/mutation_router_1.00.dat};
    \end{axis}
    \end{tikzpicture}}
    \caption{Tukey boxplots of the generation times of Firewall (top row) and Router (bottom row) versus the crossover rate ($P_c$) and mutation rate ($P_m$).  All experiments were repeated 100 times.}
    \label{fig:rates}
\end{figure}

We demonstrate GP4P4 on the 7 small network functions given in Table \ref{tab:nf}. The experiments were run on an Intel Xeon CPU E5-2690 running Ubuntu 14.04.6 LTS (kernel version 3.13.0-151).

As can be seen in Figure~\ref{fig:gen_time}, GP4P4 can generate each of the 7 network functions within 1.5 minutes. Even for the most difficult function (Router), a valid solution is usually found within 1 minute. The worst-case generation time was around 67 seconds. As network functions do not constantly need to be regenerated, this is well within acceptable limits. In fact, GP4P4 enables networks to almost immediately react to changing requirements from users or network operators, as the network can generate and install a completely new P4 program within minutes.

Next, we consider the effect of changing different parameters on the program generation time. In general, there does not seem to be a clear-cut rule for the optimal setting for \emph{all} network functions. However, in all our experiments, as long as crossover was enabled, a program could still be generated within 8 minutes at worst, suggesting that it is still possible to achieve reasonable generation times even with non-optimal parameters.

Figure~\ref{fig:parameters} shows the generation time of Firewall and Router versus the population size, tournament size ratio, and maximum initial program length. We chose to illustrate these network functions, because they have respectively the lowest and highest generation times, and thus presumably are respectively the easiest and most difficult to generate. The parameters were selected due to their impact on both functions. 

For the 4 network functions with the lowest generation times, a population size of around 1000 seems to be near-optimal. For the other network functions, a population size of 3000 gives the best results. As we want to prioritize the generation time of more difficult functions, 3000 seems to be a good choice for the population size.

For some functions, a low tournament size ratio of at most $0.1$ results in both lower generation times and generation time variance. A lower tournament size ratio allows more sub-optimal programs to evolve. Presumably, this helps increase the number of possibilities GP4P4 considers, which allows it to find valid programs more quickly.

For all network functions, limiting the maximum initial program length, max\_len, to 10 significantly improved generation times. Presumably, this is because the network functions we tested are quite small and do not require many lines of code. Alternatively, it might help early programs to satisfy one specific (implicit) behavioral rule, after which these programs can be ``merged'' using crossover in later generations.

Figure~\ref{fig:rates} shows the impact of crossover and mutation rates on the program generation time. For all network functions except Firewall and PAT, introducing crossover significantly decreases generation times as well as generation time variance. The mutation rate only has a clear impact on the generation time of the Router network function. However, as it decreased both the average generation time and the generation time variance of Router, and does not significantly increase the generation time of other network functions, mutation is clearly worthwhile to include in GP4P4.

\section{Conclusion}
\label{section:conclusion}
The size and complexity of networks has grown formidably, making managing and programming them a daunting task. In this paper, we provide a first step towards automating this process. Our proposed framework, called GP4P4, uses Linear Genetic Programming techniques to automatically generate and evolve a population of P4 network programs. GP4P4 evaluates these programs by simulating a P4 switch and generating a synthetic trace of network packets tailored towards effectively evaluating a specific rule-set. This not only reduces the computation time significantly, but also allows GP4P4 to generate P4 programs without relying on any external switches or network traces. 

Our experiments show that GP4P4 can generate P4 programs within minutes. Although GP4P4 is currently applied to simple behavioral rules, we believe it is an important first step towards a future of \emph{self-programming} networks: networks that can fully program and adapt themselves to their current goals and circumstances with minimal intervention by network operators.

\bibliographystyle{IEEEtran}
\bibliography{bibliography}

\begin{thebibliography}{10}
\providecommand{\url}[1]{#1}
\csname url@samestyle\endcsname
\providecommand{\newblock}{\relax}
\providecommand{\bibinfo}[2]{#2}
\providecommand{\BIBentrySTDinterwordspacing}{\spaceskip=0pt\relax}
\providecommand{\BIBentryALTinterwordstretchfactor}{4}
\providecommand{\BIBentryALTinterwordspacing}{\spaceskip=\fontdimen2\font plus
\BIBentryALTinterwordstretchfactor\fontdimen3\font minus
  \fontdimen4\font\relax}
\providecommand{\BIBforeignlanguage}[2]{{%
\expandafter\ifx\csname l@#1\endcsname\relax
\typeout{** WARNING: IEEEtran.bst: No hyphenation pattern has been}%
\typeout{** loaded for the language `#1'. Using the pattern for}%
\typeout{** the default language instead.}%
\else
\language=\csname l@#1\endcsname
\fi
#2}}
\providecommand{\BIBdecl}{\relax}
\BIBdecl

\bibitem{kellerer2019adaptable}
W.~Kellerer, P.~Kalmbach, A.~Blenk, A.~Basta, M.~Reisslein, and S.~Schmid,
  ``Adaptable and data-driven softwarized networks: Review, opportunities, and
  challenges,'' \emph{Proceedings of the IEEE}, vol. 107, no.~4, pp. 711--731,
  2019.

\bibitem{boutaba2018comprehensive}
R.~Boutaba, M.~A. Salahuddin, N.~Limam, S.~Ayoubi, N.~Shahriar,
  F.~Estrada-Solano, and O.~M. Caicedo, ``A comprehensive survey on machine
  learning for networking: evolution, applications and research
  opportunities,'' \emph{Journal of Internet Services and Applications},
  vol.~9, no.~1, p.~16, 2018.

\bibitem{bosshart2014p4}
P.~Bosshart, D.~Daly, G.~Gibb, M.~Izzard, N.~McKeown, J.~Rexford,
  C.~Schlesinger, D.~Talayco, A.~Vahdat, G.~Varghese \emph{et~al.}, ``P4:
  Programming protocol-independent packet processors,'' \emph{ACM SIGCOMM
  Computer Communication Review}, vol.~44, no.~3, pp. 87--95, 2014.

\bibitem{P414lang2018}
{The P4 Language Consortium}, ``P4$_{14}$ language specification,''
  \url{https://p4lang.github.io/p4-spec/p4-14/v1.0.5/tex/p4.pdf}, 2018,
  [Online; accessed 16-May-2019].

\bibitem{P416lang2017}
------, ``P4$_{16}$ language specification,''
  \url{https://p4.org/p4-spec/docs/P4-16-v1.0.0-spec.html}, 2017, [Online;
  accessed 13-January-2019].

\bibitem{Riftadi2019}
M.~Riftadi and F.~A. Kuipers, ``{P4I/O:} {Intent-Based} networking with {P4},''
  in \emph{2019 2nd International Workshop on Emerging Trends in Softwarized
  Networks (ETSN 2019 at NetSoft)}, Paris, France, Jun. 2019.

\bibitem{xie2018survey}
J.~Xie, F.~R. Yu, T.~Huang, R.~Xie, J.~Liu, C.~Wang, and Y.~Liu, ``A survey of
  machine learning techniques applied to software defined networking (sdn):
  Research issues and challenges,'' \emph{IEEE Communications Surveys \&
  Tutorials}, vol.~21, no.~1, pp. 393--430, 2018.

\bibitem{8262813}
Y.~{Geng}, S.~{Liu}, F.~{Wang}, Z.~{Yin}, B.~{Prabhakar}, and M.~{Rosenblum},
  ``Self-programming networks: Architecture and algorithms,'' in \emph{2017
  55th Annual Allerton Conference on Communication, Control, and Computing
  (Allerton)}, Oct 2017, pp. 745--752.

\bibitem{geng2019simon}
Y.~Geng, S.~Liu, Z.~Yin, A.~Naik, B.~Prabhakar, M.~Rosenblum, and A.~Vahdat,
  ``Simon: A simple and scalable method for sensing, inference and measurement
  in data center networks,'' in \emph{16th USENIX Symposium on Networked
  Systems Design and Implementation (NSDI 19)}, 2019, pp. 549--564.

\bibitem{yang2018empowering}
T.~Yang, L.~Wang, Y.~Shen, M.~Shahzad, Q.~Huang, X.~Jiang, K.~Tan, and X.~Li,
  ``Empowering sketches with machine learning for network measurements,'' in
  \emph{Proceedings of the 2018 Workshop on Network Meets AI \& ML}.\hskip 1em
  plus 0.5em minus 0.4em\relax ACM, 2018, pp. 15--20.

\bibitem{feamster2017and}
N.~Feamster and J.~Rexford, ``Why (and how) networks should run themselves,''
  \emph{arXiv preprint arXiv:1710.11583}, 2017.

\bibitem{juniper}
{Juniper Networks}, ``The self-driving network,'' March 2017, white paper.

\bibitem{kalmbach2018empowering}
P.~Kalmbach, J.~Zerwas, P.~Babarczi, A.~Blenk, W.~Kellerer, and S.~Schmid,
  ``Empowering self-driving networks,'' in \emph{Proceedings of the Afternoon
  Workshop on Self-Driving Networks}.\hskip 1em plus 0.5em minus 0.4em\relax
  ACM, 2018, pp. 8--14.

\bibitem{yaqoob2018analyzing}
T.~Yaqoob, M.~Usama, J.~Qadir, and G.~Tyson, ``On analyzing self-driving
  networks: A systems thinking approach,'' in \emph{Proceedings of the
  Afternoon Workshop on Self-Driving Networks}.\hskip 1em plus 0.5em minus
  0.4em\relax ACM, 2018, pp. 1--7.

\bibitem{Brameier2010}
M.~F. Brameier and W.~Banzhaf, \emph{Linear Genetic Programming}, 1st~ed.\hskip
  1em plus 0.5em minus 0.4em\relax Springer Publishing Company, Incorporated,
  2010.

\end{thebibliography}
\end{document}